\documentclass[showpacs,preprintnumbers,amsmath,amssymb]{revtex4}
\usepackage{graphicx}% Include figure files
\usepackage{dcolumn}% Align table columns on decimal point
\usepackage{bm}% bold math
\begin{document}
\title{Planck-scale effect through a new MDR}

% Force line breaks with \\
\author{Sohan Kumar Jha}
\affiliation{Chandernagore College, Chandernagore, Hooghly, West
Bengal, India}
\author{Himangshu Barman, Anisur Rahaman}
\email{anisur.rahman@saha.ac.in;
 manisurn@gmail.com (Corresponding Author)}
\affiliation{Hooghly Mohsin College, Chinsurah, Hooghly - 712101,
West Bengal, India}

\date{\today}% It is always \today, today,

\begin{abstract}
\begin{center}
Abstract
\end{center}
In order to the expected Planck-scale correction in the physical
systems we have put forwarded a novel modified dispersion relation
(MDR). It has a generalized structure. A specific choice of the
function used in the construction of this MDR,  it has Lorentz
invariance. A toy model like relativistic harmonic oscillator has
been studied to get the necessary  Planck-scale correction. It has
been found that each laves of harmonic oscillator acquires
Planck-scale correction and the result agrees with negligible
deviation with the result obtained for this system for the same
purpose using generalized uncertainty relation (GUP). The
relativistic Hydrogen atom problem has also been studied with this
MDR and it is found that like harmonic oscillator each energy
level of the Hydrogen atom too has got Planck-scale correction.
\end{abstract}
%\pacs{11.15Tk, 04.60Kz}
 \maketitle
\section{{\bf Introduction}}
The study of probable scenario of different physical systems in
the vicinity of Planck-scale is of huge interest since the
strength of the gravitational interaction in that scale becomes
comparable to the strength of the electromagnetic interaction. So,
in the vicinity of Planck-scale, it necessities to take into
account the effect of (quantum) gravity . But straightforward
quantization of gravity and its direct insertion into the physical
systems is still lying at far reaching stage. Therefore, the
indirect way of including the quantum gravity effect has been
receiving much attention over the years through the use of novel
ideas like generalized Heisenberg uncertainty principle and
modified dispersion relation (MDR), as these two important ideas
have been playing remarkable role towards providing necessary
Planck-scale corrections into the physical systems in indirect
manner \cite{GROSS, DAMIT, TYO, KKON, CRO, DOGL, PERE, ASHTE,
KEMPF, SUN}. This issue in the context of black body radiation was
fond to be addressed in the articles \cite{MAN, NIEM, LUB, SDAS}.
To make statistical mechanics compatible to incorporate quantum
gravity effect, formal development of it along with few
applications in the thermodynamical systems have been explored in
the articles \cite{SAH0, FIT, PPSM, PP0, PP1}. Some experimental
result also have been found to be reported to explain with the use
of this type of generalization (modification)\cite{AMELI, AMELI1}.

In the articles \cite{MAN, NIEM, LUB, SDAS, SAH0, FIT, PP0, PP1,
PP2, KIM, ALI, HOMA, HOMA1} the concept  of GUP have been
exercised extensively to incorporate Planck-scale effect into
different physical systems and the  concept of MDR  has been
extended for the same purpose in the articles \cite{NOZMDR,
BARUNMDR, ALIMDR, NOZMDR1, LINGMDR, SDASMDR, PPMDR}. Although
these two novel concepts (GUP and MDR) have been exercised in
independent manner to serve essentially the same purpose in
different physical systems, it is not difficult to understand
conceptually that MDR and GUP are complementary to each other,
however it is fair to admit that there is no established direct
one to one correspondence between these two in general. It depends
on the choice of generalization of the uncertainty relation or on
the choice of modification of the dispersion relation. We should
mention at this stage that an attempt to establish the conceptual
connection between GUP and MDR is carried out recently in the
article \cite{HFELD}. There is indeed a special instance where a
precise MDR has been proposed for a specific GUP \cite{BRM}.

Although these two substantial as well as potentially effective
ideas got strong initial support from string theory \cite{DAMIT,
KKON, PKSTRING, MTSTRING, MAGSTRING, MAGSTRING1, LJSTRING}, these
two ideas were also supported immensely by the other quantum
gravity candidates like loop quantum gravity \cite{ASHTE, NOZLOOP}
and space-time non-commutativity \cite{AR, AR1, ASU, KNOZ}. In
these two modifications (though in principle should considered to
be merged into one) Lorentz symmetry was needed to be ignored and
that necessarily led to open up a new idea namely  deformed
special relativity (DSR) \cite{SANG}. So a natural question may
arise whether this modification can be made maintaining Lorentz
symmetry. It is true that there are some experimental signature
which was explained %skillfully
 inviting the Lorentz violation in
an essential way \cite{GACNS}, but it would be nicer indeed if it
could be explained maintaining the Lorentz symmetry since
violation of Lorentz symmetry invites unwanted problems to a great
extent because this symmetry is deeply rooted both in field theory
as well as in the general theory of relativity. The articles
\cite{GAC, GAC1, GAC2} also shows a possibility of framing MDR in
a Lorentz symmetric manner where deformation of $d'$ Alembertian
has been exercised. In this article, we, therefore, introduce a
generalized modified dispersion relation in such a way that it can
be used in Lorentz invariant manner as well as with out
maintaining that invariance. In this context, we should mention
that in the articles \cite{VEG1, VEG2}, relativistic quantum
mechanical systems without quantum gravity correction have been
studied and at the same time the important article \cite{VEG}, is
of worth mentionable where possibility of incorporating invariant
quantum gravity effect at the vicinity of Planck-scale has been
discussed in detail.

So it would be beneficial to designed a generalized MDR in such a
fashion  so that it can meet both the purpose: Lorentz invariance
as well as Lorentz non-invariance extension of any physical system
towards incorporating the Planck-scale effect of any physical
system. With the MDR proposed here we have studied the toy model
like relativistic one dimensional harmonic oscillator and a three
dimensional physical system like relativistic Hydrogen atom to
include the Planck-scale correction into their energy spectrum.
Recently, these two systems has been studied in \cite{PSR1, PSR2}
for the same purpose with the use of GUP.

The article is organized in the following manner. In Sec. I, we
introduce a new generalized  dispersion relation. Sec. II is
devoted to get the Planck-scale correction of relativistic
harmonic oscillator using rasing and lowering operator. We have
shown in Sec III that the same correction can be obtained if we
use differential form of momentum operator. Sec. IV contains a
discussion of obtaining Planck-scale correction of the
relativistic Hydrogen atom directly from the
Schr$\ddot{o}$dinger's equation. In Sec. V,  the correction is
computed using the average value of different powers of momentum.
Sec. VI contains a brief discussion and conclusion.
\section{{\bf Formulation of new dispersion
relation to get the Planck-scale correctiom}} The generalized MDR
which we are going to formulate in order to incorporate quantum
gravity effect can be casted  in Lorentz invariant as well as
Lorentz non-invariant manner. So if the extension with this MDR is
carried in a Lorentz symmetric manner one need not be worried
about the search of precise DSR corresponding to this MDR. The
explicit expression of this generalized MDR is
\begin{equation}
{\bf P}^\mu {\bf P}_\mu = m_0^2+\sum_k f_k(p^2-E^2)
 ^k.\label{MDRLNS}
\end{equation}
Here $f_k = \frac{\alpha_k}{(m_P)^k}g(\frac{p}{E})$. In general,
any Lorentz non-covariant structure of the function
$g(\frac{p}{E})$ breaks the Lorentz symmetry. However any $p$
independent or Lorentz covariant
  structure of
$g(\frac{p}{E})$ preserves Lorentz invariance.  So this new
generalized MDR  can be used in any physical system to incorporate
Planck-scale correction in significant manner. We have chosen the
simplest possible Lorentz covariant structure of the function
$g(\frac{p}{E})$ , as  $g(\frac{p}{E}) =1$, which makes $f_k =
\frac{\alpha_k}{(m_P)^k}$. So the MDR with which we will start our
investigation is
\begin{equation}
{\bf P}^\mu {\bf P}_\mu =
m_0^2+\sum_k\frac{\alpha_k}{(m_P)^k}({\bf P}^\mu
 {\bf P}_\mu)^k.\label {MDR0}
\end{equation}
Here $m_0$ is the rest mass of the particle considered for study,
$k$ is an integer that runs from k = 2 to any desired order N, $P
= (E, \vec{p})$ and $p = |\vec{p}|$. The parameter $\alpha_k$
represents arbitrary N-2 number of free parameters which can be
fixed comparing the result obtained using this MDR with the
experimental result or with the results obtained earlier through
other reliable concept like GUP. The Planck-mass is characterized
by the symbol $m_p$ which is given by $m_P = \sqrt{\frac{\hbar
c}{G}}$. Note that, reparametrization invariance in the action of
a system can be maintained with this  MDR (\ref{MDR0}) having
manifestly lorentz covariance,  since under $p\rightarrow -p$ this
MDR remains unchanged. Of course, we can define  (\ref{MDR0}) in
terms of Planck-length $l_P$ in the following way
\begin{equation}
{\bf P}^\mu {\bf P}_\mu = m_0^2+\sum_k\eta_k (l_P)^k({\bf P}^\mu
 {\bf P}_\mu)^k, \label {MDR1}
\end{equation}
where $\eta_k$ represents $N-2$ number of  parameters certainly
different from ${\alpha_k}$ and the Planck-length $l_P =
\sqrt{\frac{\hbar}{Gc^3}}$. In general, there are infinite number
of parameters and numerical computation is possible with an
arbitrary large numbers of such parameters, although analytical
calculation will not always be possible with a desired numbers of
terms. In practice however we need not keep all these parameter to
get the desired accuracy.

The symbol $c$ stands for velocity of light in vacuum in both the
cases. We will choose natural unit $\hbar=1$ and $c=1$. To carry
out our analytical investigations on relativistic harmonic
oscillator and Hydrogen atom with this generalized  MDR we will
kept ourselves limited to $k=2$. In due course, we will find that
with the terms available for the the choice $k=2$, the analytical
computation towards Planck-scale correction for  the said systems
is tractable and it has good agreement with the result of the
system studied earlier using the concept of GUP to incorporate the
Planck-scale effect \cite{KEMPF}. The MDR with $k=2$ reads
\begin{equation}
E^2=p^2+m_0^2+\frac{\alpha_2}{(m_P)^2}(p^2-E^2)^2. \label {MDR2}
\end{equation}
A careful look reveals  that this MDR resembles the deformation of
the $d'$ Alembertian $(\Box)= \partial_\mu\partial^\mu$ as it has
been found in \cite{GAC, GAC1, GAC2}. To be more precise: to
include the dynamics of a physical system for any energy scale the
deformed $d'$ Alembertian would be some desired function of the
usual
 $d'$ Alembertian $(f(\Box))$ having some constants that has to be fixed
with the experimental result.  So to get Plank-scale effect it is
to be considered that the dynamics below the Plank scale will be
governed with the usual $d'$ Alembertian  and the dynamics in the
vicinity will be governed by the usual $d'$ Alembertian  along
with deformation part of the $d'$ Alembertian in a judicious
manner to get necessary agreement with the experimental result (if
or when available). Since both the has bears the Lorentz symmetry
the frame independence of the physical result will be protected.
Of course velocity of light will remain as an invariant quantity.

 The hitherto available literatures related MDR show that the
MDRs contain the expression of $E$ as a function of $P$. This new
MDR is not an exception to that, but it is true that the nature of
the function is little generalized. The modification which is made
here is only the enforcement of a physical symmetry which is none
other than the celebrated Lorentz invariance which was not there
in the MDR designed earlier. We have kept this  point of view that
in the vicinity of the Planck-scale violation of Lorentz
invariance may occur but it can not be a basic criteria. On the
other hand it would be admitted from all corners that Lorentz
covariant structure is advantageous over Lorentz non-covariance in
many respect because the basic foundation of quantum field theory
and general theory of relativity is deeply rooted to the Poincar'e
symmetry.  The article \cite{SMO} is an excellent example in this
direction Therefore, this new MDR will certainly add new light in
the formal development of MDR related theories. The novelty of
this MDR is the welcome entry of the Lorentz symmetry and its
capability to render Planck-scale correction in a frame
independent manner which is lacking in the construction of MDR in
the available literature dealing with Planck-scale correction with
MDR.

Taking  the square root of the above expression we get
\begin{equation}
E=\sqrt{p^2+m_0^2+\frac{\alpha_2}{(m_P)^2}(p^2-E^2)^2}.
\label{EMDR2}
\end{equation}
Equation (\ref{EMDR2}) on binomial expansion results
\begin{eqnarray}
E&=&\frac{p^2}{2m_0}+m_0+\frac{\alpha_2
E_n^4}{2m_0(m_P)^2}-p^2(\frac{\alpha_2
E_n^4}{4m_0^3(m_P)^2}+\frac{\alpha_2 E_n^2}{m_0(m_P)^2})\nonumber\\
&+&p^4(\frac{\alpha_2}{2m_0(m_P)^2}+\frac{\alpha_2
E_n^2}{2m_0^3(m_P)^2}-\frac{1}{8(m_0)^3}
+\frac{3\alpha_2 E_n^4}{16m_0^5(m_P)^2})\nonumber\\
&+&p^6(-\frac{\alpha_2}{4m_0^3(m_P)^2}+\frac{1}{16(m_0)^5}-\frac{3\alpha_2
E_n^2}{8m_0^5(m_P)^2}-\frac{5\alpha_2 E_n^4}{32m_0^7(m_P)^2}).
\end{eqnarray}
In the above expansion we have neglected terms containing higher
order in $\alpha_2$ and retained the terms up to sixth order in p.
Thus our modified Hamiltonian with this settings reads
\begin{eqnarray}
H&=&\frac{p^2}{2m_0}+V+m_0+\frac{\alpha
E_n^4}{2m_0(m_P)^2}-p^2(\frac{\alpha
E_n^4}{4m_0^3(m_P)^2}+\frac{\alpha E_n^2}{m_0(m_P)^2})\nonumber\\
&+&p^4(\frac{\alpha}{2m_0(m_P)^2}+\frac{\alpha
E_n^2}{2m_0^3(m_P)^2}-\frac{1}{8(m_0)^3}
+\frac{3\alpha E_n^4}{16m_0^5(m_P)^2})\nonumber\\
&+&p^6(-\frac{\alpha}{4m_0^3(m_P)^2}+\frac{1}{16(m_0)^5}-\frac{3\alpha
E_n^2}{8m_0^5(m_P)^2}-\frac{5\alpha E_n^4}{32m_0^7(m_P)^2}).
\label{HAM}
\end{eqnarray}
In the above expression we have replaced $\alpha_2$ by $\alpha$
since there is no other $\alpha$'s in the body of the article.
Here $V$ stands for the potential energy inserted by hand. To
avoid confusion we would like to mention that $[q,~p] = i\hbar$
and for modified position and momentum pair $(q_m,~p_m)$ the
canonical Poission's bracket will take a modified form and that
will lead to a modified uncertainty relation having a generalized
form $[q_m,~p_m] = i\hbar f(p, E)$. The precise form of the
function $f(p, E)$ will certainly depend on the nature of the
choice of the MDR. Now using this Hamiltonian we will proceed to
calculate modified eigenvalues for the relativistic harmonic
oscillator and Hydrogen atom. The modified eigenvalues are in
general given by
\begin{eqnarray}
{E_{n}}^{(m)}&=& E_n+m_0+\frac{\alpha
E_n^4}{2m_0(m_P)^2}-<p^2>(\frac{\alpha
E_n^4}{4m_0^3(m_P)^2}+\frac{\alpha E_n^2}{m_0(m_P)^2})\nonumber\\
&+&<p^4>(\frac{\alpha}{2m_0(m_P)^2}+\frac{\alpha
E_n^2}{2m_0^3(m_P)^2}-\frac{1}{8(m_0)^3}
+\frac{3\alpha E_n^4}{16m_0^5(m_P)^2})\nonumber\\
&+&
<p^6>(-\frac{\alpha}{4m_0^3(m_P)^2}+\frac{1}{16(m_0)^5}-\frac{3\alpha
E_n^2}{8m_0^5(m_P)^2}-\frac{5\alpha E_n^4}{32m_0^7(m_P)^2}),
\label{EMOD}
\end{eqnarray}
where ${E_{n}}^{(m)}$ and $E_n$ are modified and unmodified
eigenvalues respectively. This generalized expression shows that
we have to evaluate the expectation values up to sixth power of
momentum. To be precise we need the expressions of $<p^2>$,
$<p^4>$, and $<p^6>$ of the system that would considered under
investigation.
\section{{\bf Relativistic Harmonic Oscillator with  MDR
using raising and lowering representation of momentum operator}}
Let us first proceed to calculate the energy eigenvalues using
raising and lowering representation of the momentum. It is known
that the raising and lowering operators are respectively given by
\begin{eqnarray}
a&=&\frac{1}{\sqrt{2m_0\omega\hbar}}(m_0\omega x+ip),
\\a^\dagger&=&\frac{1}{\sqrt{2m_0\omega\hbar}}(m_0\omega x-ip).
\end{eqnarray}
For the sake of convenience we will not set $\hbar =1$ but $c=1$
will be maintained from the starting point. However the final
expression will be presented with $\hbar =1$. If $a$ and $a^+$ are
operated separately on the $n^{th}$ eigen state $|n>$ we get
\begin{eqnarray}
a|n>&=&\sqrt{n}|n-1>,
\\a^\dagger|n>&=&\sqrt{n+1}|n+1>.
\end{eqnarray}
The momentum operator in terms of raising and lowering operator
can be expressed as
\begin{equation}
p=i\sqrt{\frac{m_0\omega\hbar}{2}}(a^\dagger-a)
\end{equation}
Consecutive  operation of $p$  on the $n^{th}$ eigen state for
two, four and six times  results the following
\begin{eqnarray}
p^2|n>&=&-\frac{m_0\omega\hbar}{2}[\sqrt{n+1}\sqrt{n+2}|n+2>-(2n+1)|n>+\sqrt{n}\sqrt{n-1}|n-2>],\\
p^4|n>&=&(\frac{m_0\omega\hbar}{2})^2[\sqrt{n+1}\sqrt{n+2}\sqrt{n+3}\sqrt{n+4}|n+4>
\\\nonumber&-&\sqrt{n+1}\sqrt{n+2}(4n+6)|n+2>\\\nonumber&+&(6n^2+6n+3)|n>
-\sqrt{n}\sqrt{n-1}(4n-2)|n-2>\\\nonumber&+&\sqrt{n}\sqrt{n-1}\sqrt{n-2}\sqrt{n-3}|n-4>],\\
p^6|n>&=&-(\frac{m_0\omega\hbar}{2})^3[\sqrt{n+1}\sqrt{n+2}\sqrt{n+3}\sqrt{n+4}\sqrt{n+5}\sqrt{n+6}
|n+6>\\\nonumber&-&\sqrt{n+1}\sqrt{n+2}\sqrt{n+3}\sqrt{n+4}(6n+15)|n+4>\\\nonumber
&+&\sqrt{n+1}\sqrt{n+2}(15n^2+45n+45)|n+2>\\\nonumber&-&(20n^3+30n^2+40n+15)|n>
\\\nonumber&+&\sqrt{n}\sqrt{n-1}(15n^2-15n+15)|n-2>
\\\nonumber&-&\sqrt{n}\sqrt{n-1}\sqrt{n-2}\sqrt{n-3}(6n-9)|n-4>
\\\nonumber&+&\sqrt{n}\sqrt{n-1}\sqrt{n-2}\sqrt{n-3}\sqrt{n-4}\sqrt{n-5}|n-6>].
\end{eqnarray}
If we now take the inner product of the above expression with
$|n>$ we will get the required expectation values of $p^2$, $p^4$
and $p^6$. The precise expression of the expectation values are
\begin{eqnarray}
<p^2>&=&\frac{m_0\omega\hbar}{2}(2n+1),\\
<p^4>&=&(\frac{m_0\omega\hbar}{2})^2(6n^2+6n+3),\\
<p^6>&=&(\frac{m_0\omega\hbar}{2})^3(20n^3+30n^2+40n+15),
\end{eqnarray}
which leads us to get the modified eigenvalues with this Lorentz
symmetric modified dispersion relation
\begin{eqnarray}
{E_{n}}^{(mho)}&=&E_n+m_0+\frac{\alpha
E_n^4}{2m_0(m_P)^2}-\frac{m_0\omega\hbar}{2}(2n+1)(\frac{\alpha
E_n^4}{4m_0^3(m_P)^2}+\frac{\alpha E_n^2}{m_0(m_P)^2})\nonumber \\
&+&(\frac{m_0\omega\hbar}{2})^2(6n^2+6n+3)(\frac{\alpha}{2m_0(m_P)^2}+\frac{\alpha
E_n^2}{2m_0^3(m_P)^2}-\frac{1}{8(m_0)^3}+\frac{3\alpha
E_n^4}{16m_0^5(m_P)^2})\nonumber \\
&+&(\frac{m_0\omega\hbar}{2})^3 \times (20n^3+30n^2+40n+15)\nonumber \\
&(&-\frac{\alpha}{4m_0^3(m_P)^2}+\frac{1}{16(m_0)^5}-\frac{3\alpha
E_n^2}{8m_0^5(m_P)^2}-\frac{5\alpha E_n^4}{32m_0^7(m_P)^2}).
\label{LSHO}
\end{eqnarray}
Equation (\ref{LSHO}) reveals that each levels of hydrogen atom
acquires Planck-scale correction which was also found in the
articles \cite{PSR1, PSR2} where concept of GUP was employed to
incorporate quantum gravity effect. This result is amenable to
compare with the correction obtained in \cite{KEMPF}, using the
concept of GUP for incorporation of quantum gravity correction. We
will now turn to calculate the modified energy eigenvalues using
the differential form of the momentum to get it confirmed whether
these two agrees with each other.
\subsection{Calculation using differential form of momentum operator}
We know that momentum operator in differential form with the
coordinate representation can is written down as
\begin{equation}
p =-i\hbar\frac{\partial}{\partial x},~~~~i=\sqrt{-1},
\end{equation}
and the $n^{th}$ state eigenfunction of a harmonic oscillator is
known to be
\begin{equation}
\psi_n=\sqrt{\frac{a}{\sqrt{\pi}2^nn!}}H_n(ax)e^{\frac{-a^2x^2}{2}},
a=\sqrt{\frac{m_0\omega}{\hbar}},
\end{equation}
where $H_n(ax)$, represents the Hermite polynomial of order n.
Bringing the  momentum operator in repeated action for two, four
and six times and multiplying the obtained result by $\psi_n$ and
hence integrating within the limit $-\infty$ to $\infty$ we  get
the expectation values of $p^2$, $p^4$ and $p^6$:
\begin{eqnarray}
<p^2>=&-&\hbar^2\frac{a}{\sqrt{\pi}2^nn!}[\int_{-\infty}^{\infty}(a^4x^2-a^2)e^{-a^2x^2}
H_n(ax)H_n(ax)dx\\\nonumber&-&4na^3\int_{-\infty}^{\infty}xe^{a^2x^2}H_{n-1}(ax)H_n(ax)dx\\
\nonumber&+&4n(n-1)a^2\int_{-\infty}^{\infty}e^{a^2x^2}H_{n-2}(ax)H_n(ax)dx]\\
&=&\frac{m_0\omega\hbar}{2}(2n+1),
\end{eqnarray}
\begin{eqnarray}
 <p^4>=&
&\hbar^4\frac{a}{\sqrt{\pi}2^nn!}[16n(n-1)(n-2)(n-3)a^4\int_{-\infty}^{\infty}e^{a^2x^2}H_{n-4}(ax)H_n(ax)dx
\nonumber \\
&-&32n(n-1)(n-2)a^5\int_{-\infty}^{\infty}xe^{a^2x^2}H_{n-3}(ax)H_n(ax)dx\nonumber \\
&+&24n(n-1)\int_{-\infty}^{\infty}(a^6x^2-a^4)e^{a^2x^2}H_{n-2}(ax)H_n(ax)dx
\nonumber \\
&+&\int_{-\infty}^{\infty}(24na^5x-8na^7x^3)e^{a^2x^2}H_{n-1}(ax)H_n(ax)dx
\nonumber \\&+&\int_{-\infty}^{\infty}(3a^4-6a^6x^2+a^8x^4)e^{a^2x^2}H_{n}(ax)H_n(ax)dx]\\
&=&(\frac{m_0\omega\hbar}{2})^2(6n^2+6n+3),
\end{eqnarray}
\begin{eqnarray}
<p^6>=&-&\hbar^6\frac{a}{\sqrt{\pi}2^nn!}[64n(n-1)(n-2)(n-3)(n-4)(n-5)a^6
\int_{-\infty}^{\infty}e^{a^2x^2}H_{n-6}(ax)H_n(ax)dx \nonumber \\
&-&192n(n-1)(n-2)(n-3)(n-4)a^7\int_{-\infty}^{\infty}xe^{a^2x^2}H_{n-5}(ax)H_n(ax)dx
\nonumber \\
&+&240n(n-1)(n-2)(n-3)\int_{-\infty}^{\infty}(a^8x^2-a^6)e^{a^2x^2}H_{n-4}(ax)H_n(ax)dx
\nonumber \\
&+&n(n-1)(n-2)\int_{-\infty}^{\infty}(480a^7x-160a^9x^3)e^{a^2x^2}H_{n-3}(ax)H_n(ax)dx
\nonumber \\
&+&n(n-1)\int_{-\infty}^{\infty}(180a^6-360a^8x^2+60a^10x^4)e^{a^2x^2}H_{n-2}(ax)H_n(ax)dx
\nonumber \\
&+&n\int_{-\infty}^{\infty}(-180a^7x+120a^9x^3-12a^11x^5)e^{a^2x^2}H_{n-1}(ax)H_n(ax)dx
\nonumber \\
&+&\int_{-\infty}^{\infty}(45a^8x^2-15a^6-15a^10x^4+a^12x^6)e^{a^2x^2}H_{n}(ax)H_n(ax)dx
\\
&=&(\frac{m_0\omega\hbar}{2})^3(20n^3+30n^2+40n+15).
\end{eqnarray}
In the above computation we have used the following recurrence
relations.
\begin{eqnarray}
H_{n+1}(x)=2xH_n(x)-2nH_{n-1}(x),~~~
\frac{dH_n(x)}{dx}=2nH_{n-1}(x).
\end{eqnarray}
The the standard integrals
\begin{eqnarray}
\int_{-\infty}^{\infty}e^{x^2}H_{m}(x)H_n(x)dx&=&2^n\sqrt{\pi}n!\delta_{mn},\\
\int_{-\infty}^{\infty}xe^{x^2}H_{m}(x)H_n(x)dx&=&2^{n-1}\sqrt{\pi}n!\delta_{m,n-1}+2^{n}
\sqrt{\pi}(n+1)!\delta_{m,n+1},\\
\int_{-\infty}^{\infty}x^2e^{x^2}H_{m}(x)H_n(x)dx&=&2^{n-2}\sqrt{\pi}n!\delta_{m,n-2}+(2n+1)2^{n-1}
\sqrt{\pi}n!\delta_{m,n} +2^{n}\sqrt{\pi}(n+2)!\delta_{m,n+2},
\end{eqnarray}
are also needed for the required computation. Using the above
recurrence relations and the standard integrals the final
expression of the modified eigenvalues come out to be
\begin{eqnarray}
{E_{n}}^{(mho)}&=&E_n+m_0+\frac{\alpha
E_n^4}{2m_0(m_P)^2}-\frac{m_0\omega\hbar}{2}(2n+1)(\frac{\alpha
E_n^4}{4m_0^3(m_P)^2}+\frac{\alpha E_n^2}{m_0(m_P)^2})\nonumber \\
&+&(\frac{m_0\omega\hbar}{2})^2(6n^2+6n+3)(\frac{\alpha}{2m_0(m_P)^2}+\frac{\alpha
E_n^2}{2m_0^3(m_P)^2}-\frac{1}{8(m_0)^3}+\frac{3\alpha
E_n^4}{16m_0^5(m_P)^2})\nonumber \\
&+&(\frac{m_0\omega\hbar}{2})^3(20n^3+30n^2+40n+15)\nonumber\\
&&(-\frac{\alpha}{4m_0^3(m_P)^2}+\frac{1}{16(m_0)^5}-\frac{3\alpha
E_n^2}{8m_0^5(m_P)^2}-\frac{5\alpha E_n^4}{32m_0^7(m_P)^2}).
\label{LSHO1}
\end{eqnarray}
Note that  identical  expression for eigenvalue ${E_{n}}^{(m)}$
has come out for both the cases: when momentum is expressed in
terms of rasing and lowering operator as well as when the
differential form of the momentum operator is used for
computation. This is of course, the expected scenario. The
expression obtained in (\ref{LSHO}) and (\ref{LSHO1}) do not have
Lorentz symmetric structure although the the Lorentz covariant MDR
has been used for computation of eigenvalues, because we have not
considered the full relativistic theory. For full relativistic
Hamiltonian the result ought to be Lorentz symmetric if at all the
exact analytical extension is feasible in this situation.

Fig.1 shows a plot of the energy eigenvalues versus $n$ for
$\alpha =1.6\times 10^{-39}$. Note that energy eigenvalues is
plotted subtracting the rest mass energy from it.  The spectrum
agrees with spectrum obtained in  the article \cite{KEMPF} which
was evaluated using a quadratic GUP.  In our case modification is
resulted with the introduction of a generalized MDR. It ravels
once again that GUP and MDR essentially serve the same purpose
which is indeed the obtaining the quantum gravity correction at
the vicinity of Planck-scale. The agreement of our result with the
result obtained in the article \cite{KEMPF} is natural since the
MDR and GUP basically serves the same purpose although it is fair
to admit that there is no one to one correspondence between this
generalized MDR used here and the GUP used in \cite{KEMPF}.
%\begin{center}
\begin{figure}[!hb]
   \centering
   \vspace{2cm}
   \fbox{
   \includegraphics[width=16cm,height=10cm]{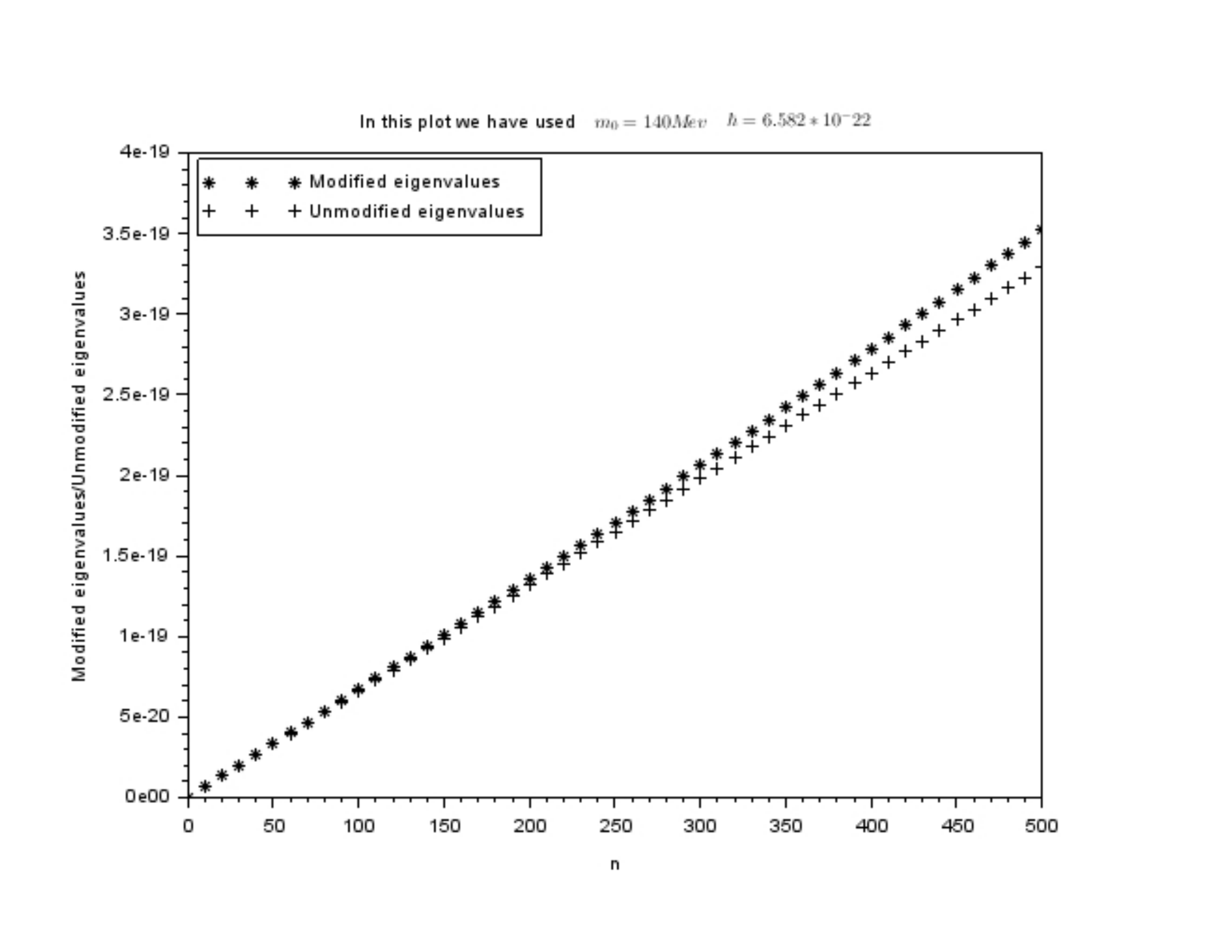}}
   \caption{A Picture}\label{fig:Growth}
\end{figure}
%\end{wrapfigure}
%\end{center}
\section{Modified eigenvalues of Hydrogen atom using
Schr$\ddot{o}$dinger's  equation} Let us now proceed to obtain
quantum gravity correction to the spectrum of Hydrogen atom. The
Schr$\ddot{o}$dinger's equation for Hydrogen atom reads
\begin{equation}
\frac{p^2}{2m_0}\psi_{nlm}-\frac{e^2}{4\pi\varepsilon_0r}\psi_{nlm}=E_n\psi_{nlm},
\end{equation}
where
\begin{equation}
\psi_{nlm}=R_{nl}Y_l^m(\theta\varphi)=\frac{u_{nl}}{r}Y_l^m(\theta\varphi),
\label{EFHA}
\end{equation}
 This gives
\begin{equation}
p^2\psi_{nlm}=2m_0[E_n+\frac{e^2}{4\pi\varepsilon_0r}]\psi_{nlm},
\end{equation}
and hence the expectation value of $p^2$ is given by
\begin{equation}
<p^2>=2m_0[E_n+\frac{e^2}{4\pi\varepsilon_0}<\frac{1}{r}>].
\end{equation}
If $p^4$ and $p^6$ on $\psi_{nlm}$  get operated on (\ref{EFHA})
we ultimately have
\begin{eqnarray}
p^4(\psi_{nlm})&=&4m_0^2E_n[E_n+\frac{e^2}{4\pi\varepsilon_0r}]\psi_{nlm}
+\frac{m_0e^2}{2\pi\varepsilon_0}[\frac{2\hbar^2Y_l^m}{r^2}\frac{dR_{nl}}{dr}
+\frac{2m_0}{r}(E_n+\frac{e^2}{4\pi\varepsilon_0r})\psi_{nlm}],\\\nonumber
p^6(\psi_{nlm})&=&8m_0^3E_n^2[E_n+\frac{e^2}{4\pi\varepsilon_0r}]\psi_{nlm}
+\frac{2m_0^2e^2E_n}{\pi\varepsilon_0}[\frac{2\hbar^2Y_l^m}{r^2}\frac{dR_{nl}}{dr}
+(\frac{2m_0E_n}{r}+\frac{m_0e^2}{2\pi\varepsilon_0r^2})\psi_{nlm}]\\\nonumber
&+&\frac{m_0^2e^4}{4\pi^2\varepsilon^2}[(\frac{m_0e^2}{2\pi\varepsilon_0r^3}\frac{2m_0E_n}{r^2})\psi_{nlm}
+\frac{4\hbar^2Y_l^m}{r^4}\frac{du_{nl}}{dr}-\frac{6\hbar^2}{r^4}\psi_{nlm}]\\\nonumber
&+&\frac{m_0e^2\hbar^2}{\pi\varepsilon_0}[\frac{l(l+1)\hbar^2Y_l^m}{r^4}\frac{dR_{nl}}{dr}
+\frac{12\hbar^2}{r^5}\psi_{nlm}+(\frac{5l(l+1)\hbar^2}{r^5}-\frac{5m_0e^2}{2\pi\varepsilon_0r^4}
-\frac{10m_0E_n}{r^3})\psi_{nlm}\\
&-&\hbar^2Y_l^m(\frac{1}{r^3}\frac{d^3u_{nl}}{dr^3}+\frac{12}{r^5}\frac{du_{nl}}{dr})].
\end{eqnarray}
These lead us to obtain the expectation value of $p^4$:
\begin{equation}
<p^4>=4m_0^2E_n^2+\frac{2m_0^2e^2E_n}{\pi\varepsilon_0}<\frac{1}{r}>
+\frac{m_0^2e^4}{4\pi^2\varepsilon_0^2}<\frac{1}{r^2}>.
\end{equation}
To get the desired result we have used the stand integral
\begin{equation}
\int_0^\infty R_{nl}\frac{dR_{nl}}{dr}dr=0.
\end{equation}
The expectation value of $p^6$ is now found out in a
straightforward manner:
\begin{eqnarray}\nonumber
<p^6>=& &8m_0^3E_n^3+\frac{6m_0^3e^2E_n^2}{\pi\varepsilon_0}<\frac{1}{r}>
+\frac{3m_0^3e^4E_n}{2\pi^2\varepsilon^2}<\frac{1}{r^2}>+[\frac{m_0^3e^6}{8\pi^3\varepsilon_0^3}
-\frac{8m_0^2E_ne^2\hbar^2}{\pi\varepsilon_0}]<\frac{1}{r^3}>\\
&-&\frac{9m_0^2e^4\hbar^2}{4\pi^2\varepsilon_0^2}<\frac{1}{r^4}>
+\frac{m_0e^2\hbar^2}{\pi\varepsilon_0}[6l(l+1)\hbar^2-12\hbar^2]<\frac{1}{r^5}>.
\end{eqnarray}
To this end, Feynmann-Hellman theorem helps a lot to get the
expressions
 of $<\frac{1}{r}>$ and $<\frac{1}{r^2}>$:
Ultimately, we see that
\begin{eqnarray}
<\frac{1}{r}>&=&\frac{1}{n^2a_0},\\
<\frac{1}{r^2}>&=&\frac{2}{n^3a_0^2(2l+1)},
\end{eqnarray}
where $a_0$ is the Bohr radius. To get the exact expression of
$<\frac{1}{r^3}>$,$<\frac{1}{r^4}>$ and $<\frac{1}{r^5}>$ Kramers'
relation \cite{GRIF} also has been employed here which is given by
\begin{equation}
\frac{s+1}{n^2}<r^s>-(2s+1)a_0<r^{s-1}>+\frac{s}{4}[(2l+1)^2-s^2]a_0^2<r^{s-2}>=0.
\end{equation}
If $s=-1$ is set in the Kramers' relation it leads to obtain the
expression of  $<\frac{1}{r^3}>$:
\begin{equation}
<\frac{1}{r^3}>=\frac{2}{a_0^3n^3l(l+1)(2l+1)},
\end{equation}
In a similar way it we put $s=-2$ in the Kramers' relation it
gives the expression of $<\frac{1}{r^4}>$
\begin{equation}
<\frac{1}{r^4}>=\frac{12}{a_0^4n^3l(l+1)(2l+3)(4l^2-1)}-\frac{4}{a_0^4n^5(2l+3)(4l^2-1)}.
\end{equation}
Finally,  putting $s=-3$ we get the expression of
$<\frac{1}{r^5}>$:
\begin{eqnarray}\nonumber
<\frac{1}{r^5}>=&
&\frac{20}{a_0^5n^3l(l^2-1)(l+2)(2l+3)(4l^2-1)}\\\nonumber
&-&\frac{20}{3a_0^5n^5(l-1)(l+2)(2l+3)(4l^2-1)}\\&-&\frac{4}{3a_0^5n^5l(l^2-1)(l+2)(2l+1)}.
\end{eqnarray}
Substituting these  in the expression for $<p^2>$, $<p^4>$ and
$<p^6>$ we land on to the following.
\begin{eqnarray}
<p^2>&=&2m_0[E_n+\frac{\hbar^2}{m_0a_0^2n^2}],
\end{eqnarray}
\begin{eqnarray}
<p^4>&=&\frac{8\hbar^4}{a_0^4n^3(2l+1)}+4m_0^2E_n^2+\frac{8m_0E_n\hbar^2}{n^2a_0^2},
\end{eqnarray}
\begin{eqnarray}
<p^6>&=&8m_0^3E_n^3+\frac{6m_0^3e^2E_n^2}{\pi\varepsilon_0}\frac{1}{n^2a_0}\nonumber \\
&+&\frac{3m_0^3e^4E_n}{2\pi^2\varepsilon^2}\frac{2}{n^3a_0^2(2l+1)}
+[\frac{m_0^3e^6}{8\pi^3\varepsilon_0^3}
-\frac{8m_0^2E_ne^2\hbar^2}{\pi\varepsilon_0}]\frac{2}{a_0^3n^3l(l+1)(2l+1)}\nonumber \\
&-&\frac{9m_0^2e^4\hbar^2}{4\pi^2\varepsilon_0^2}[\frac{12}{a_0^4n^3l(l+1)(2l+3)(4l^2-1)}
-\frac{4}{a_0^4n^5(2l+3)(4l^2-1)}]\nonumber \\
&+&\frac{m_0e^2\hbar^2}{\pi\varepsilon_0}[6l(l+1)\hbar^2-12\hbar^2][\frac{20}{a_0^5n^3l(l^2-1)(l+2)(2l+3)(4l^2-1)}\nonumber \\
&-&\frac{20}{3a_0^5n^5(l-1)(l+2)(2l+3)(4l^2-1)}-\frac{4}{3a_0^5n^5l(l^2-1)(l+2)(2l+1)}].
\end{eqnarray}
Thus the modified eigenvalues for Hydrogen atom finally come out
as
\begin{eqnarray}
{E_{n}}^{(mha)}=& & E_n+m_0+\frac{\alpha
E_n^4}{2m_0(m_P)^2}-2m_0[E_n+\frac{\hbar^2}{m_0a_0^2n^2}](\frac{\alpha
E_n^4}{4m_0^3(m_P)^2}+\frac{\alpha E_n^2}{m_0(m_P)^2})\nonumber \\
&+&[\frac{8\hbar^4}{a_0^4n^3(2l+1)}+4m_0^2E_n^2+\frac{8m_0E_n\hbar^2}{n^2a_0^2}]\nonumber
\\ & &(\frac{\alpha}{2m_0(m_P)^2}+\frac{\alpha
E_n^2}{2m_0^3(m_P)^2}-\frac{1}{8(m_0)^3} +\frac{3\alpha
E_n^4}{16m_0^5(m_P)^2})\nonumber \\
&+&[8m_0^3E_n^3+\frac{6m_0^3e^2E_n^2}{\pi\varepsilon_0}\frac{1}{n^2a_0}
+\frac{3m_0^3e^4E_n}{2\pi^2\varepsilon^2}\frac{2}{n^3a_0^2(2l+1)}\nonumber \\
&+&(\frac{m_0^3e^6}{8\pi^3\varepsilon_0^3}
-\frac{8m_0^2E_ne^2\hbar^2}{\pi\varepsilon_0})\frac{2}{a_0^3n^3l(l+1)(2l+1)}\nonumber \\
&-&\frac{9m_0^2e^4\hbar^2}{4\pi^2\varepsilon_0^2}(\frac{12}{a_0^4n^3l(l+1)(2l+3)(4l^2-1)}
-\frac{4}{a_0^4n^5(2l+3)(4l^2-1)})\nonumber \\
&+&\frac{m_0e^2\hbar^2}{\pi\varepsilon_0}(6l(l+1)\hbar^2-12\hbar^2)(\frac{20}
{a_0^5n^3l(l^2-1)(l+2)(2l+3)(4l^2-1)}\nonumber \\
&-&\frac{20}{3a_0^5n^5(l-1)(l+2)(2l+3)(4l^2-1)}-\frac{4}{3a_0^5n^5l(l^2-1)(l+2)(2l+1)})]
\nonumber \\
&&(-\frac{\alpha}{4m_0^3(m_P)^2}+\frac{1}{16(m_0)^5}-\frac{3\alpha
E_n^2}{8m_0^5(m_P)^2}-\frac{5\alpha
E_n^4}{32m_0^7(m_P)^2}).\label{LSHA}
\end{eqnarray}
Like the harmonic oscillator this expression also does not have
Lorentz invariance. The reason indeed is the same as we have
maintained in earlier when we got the modified eigenvalues of the
harmonic oscillator: the full relativistic theory in this case too
is not possible to consider for obtaining analytical computation.
What follows next is the computation of eigenvalues Hydrogen atom
calculating the average values of different  power of momentum as
required.
\subsection{Spectrum through computation of average values  of different powers of momentum}
The differential form of $p^2$ in polar coordinate is written down
as
\begin{equation}
p^2=-\frac{\hbar^2}{r^2}\frac{\partial}{\partial
r}(r^2\frac{\partial}{\partial
r})-\frac{\hbar^2}{r^2}[\frac{1}{\sin\theta}\frac{\partial}{\partial
\theta}(\sin\theta\frac{\partial}{\partial\theta})
+\frac{1}{\sin^2\theta}\frac{\partial^2}{\partial^2\varphi}].
\end{equation}
This can be rewritten as
\begin{equation}
p^2=p_r^2+\frac{L^2}{r^2},
\end{equation}
where
\begin{equation}
p_r^2=-\frac{\hbar^2}{r^2}\frac{\partial}{\partial
r}(r^2\frac{\partial}{\partial r})
\end{equation}
 and
\begin{equation}
L^2=-\hbar^2[\frac{1}{\sin\theta}\frac{\partial}{\partial
\theta}(\sin\theta\frac{\partial}{\partial\theta})+\frac{1}{\sin^2\theta}
\frac{\partial^2}{\partial^2\varphi}].
\end{equation}
The eigenfunction for Hydrogen atom is given by
\begin{equation}
\psi_{nlm}=\frac{u_{nl}(r)}{r}Y_l^m(\theta,\varphi),
\end{equation}
where $Y_l^m(\theta,\varphi)$ represent the spherical harmonics.
If we operate $p^2$ on $\psi_{nlm}$ once, twice and thrice we get
\begin{eqnarray}
p^2(\psi_{nlm})&=&-\frac{\hbar^2Y_l^m}{r}\frac{d^2u_{nl}}{dr^2}
+\frac{l(l+1)\hbar^2}{r^2}
\psi_{nlm}, \\
p^4(\psi_{nlm})&=&-\frac{m_0e^2\hbar^2}{\pi\varepsilon_0r^4}Y_l^mu_{nl}
+\frac{m_0e^2\hbar^2}
{\pi\varepsilon_0r^3}Y_l^m\frac{du_{nl}}{dr},\nonumber \\
&+&4m_0^2(\frac{e^2}{\pi\varepsilon_0})^2\frac{1}{r^3}Y_l^mu_{nl}
+\frac{4m_0^2E_n^2}{r}Y_l^mu_{nl}\nonumber \\
&+&\frac{2m_0^2e^2E_n}{\pi\varepsilon_0r^2}Y_l^mu_{nl} \\
p^6(\psi_{nlm})&=&8m_0^3E_n^2[E_n+\frac{e^2}{4\pi\varepsilon_0r}]\psi_{nlm}\nonumber\\
&+&\frac{2m_0^2e^2E_n}{\pi\varepsilon_0}[\frac{2\hbar^2}{r^2}Y_l^m\frac{dR_{nl}}{dr}
+\frac{2m_0}{r}(E_n+\frac{e^2}{4\pi\varepsilon_0r})\psi_{nlm}]\nonumber \\
&+&\frac{m_0^2e^4}{4\pi^2\varepsilon_0^2}[\frac{l(l+1)\hbar^2}{r^4}\psi_{nlm}
-\frac{2m_0}{r}(\frac{l(l+1)\hbar^2}{2m_0r^2}
-\frac{e^2}{4\pi\varepsilon_0r}-E_n)\psi_{nlm}\nonumber \\
&+&\frac{4\hbar^2}{r^4}Y_l^m\frac{du_{nl}}{dr}-\frac{6\hbar^2}{r^4}\psi_{nlm}]
+\frac{m_0e^2\hbar^2}{\pi\varepsilon_0}[\frac{l(l+1)\hbar^2}{r^4}Y_l^m\frac{dR_{nl}}{dr}
\nonumber\\
&+&\frac{12\hbar^2}{r^5}\psi_{nlm}+\frac{10m_0}{r^3}(\frac{l(l+1)\hbar^2}{2m_0r^2}
-\frac{e^2}{4\pi\varepsilon_0r}-E_n)\psi_{nlm}\nonumber \\
&-&\hbar^2Y_l^m(\frac{1}{r^3}\frac{d^3u_{nl}}{dr^3}+\frac{12}{r^5}\frac{du_{nl}}{dr})].
\end{eqnarray}
If we multiply the results by $\psi^*_{nlm}$ and then integrate
within the limit $0$ to $\infty$, $0$ to $\pi$ and $0$ to $2\pi$
for $r$,$\theta$ and $\phi$ respectively we get the expectation
value of $p^2$, $p^4$ and $p^6$ as
\begin{eqnarray}
<p^2>&=&2m_0[E_n+\frac{e^2}{4\pi\varepsilon_0}<\frac{1}{r}>]\nonumber \\
&=&2m_0[E_n+\frac{\hbar^2}{m_0a_0^2n^2}],
\end{eqnarray}
\begin{eqnarray}
<p^4>&=&4m_0^2E_n^2+\frac{2m_0^2e^2E_n}{\pi\varepsilon_0}<\frac{1}{r}>
+\frac{m_0^2e^4}{4\pi^2\varepsilon_0^2}<\frac{1}{r^2}>\nonumber \\
&=&\frac{8\hbar^4}{a_0^4n^3(2l+1)}+4m_0^2E_n^2+\frac{8m_0E_n\hbar^2}{n^2a_0^2},
\end{eqnarray}
\begin{eqnarray}
<p^6>&=&8m_0^3E_n^3+\frac{6m_0^3e^2E_n^2}{\pi\varepsilon_0}<\frac{1}{r}>
+\frac{3m_0^3e^4E_n}{2\pi^2\varepsilon^2}<\frac{1}{r^2}>
+[\frac{m_0^3e^6}{8\pi^3\varepsilon_0^3}
-\frac{8m_0^2E_ne^2\hbar^2}{\pi\varepsilon_0}]<\frac{1}{r^3}>\nonumber \\
&-&\frac{9m_0^2e^4\hbar^2}{4\pi^2\varepsilon_0^2}<\frac{1}{r^4}>
+\frac{m_0e^2\hbar^2}{\pi\varepsilon_0}[6l(l+1)\hbar^2-12\hbar^2]
<\frac{1}{r^5}>\nonumber \\
&=&8m_0^3E_n^3+\frac{6m_0^3e^2E_n^2}{\pi\varepsilon_0}\frac{1}{n^2a_0}\nonumber \\
&+&\frac{3m_0^3e^4E_n}{2\pi^2\varepsilon^2}\frac{2}{n^3a_0^2(2l+1)}
+[\frac{m_0^3e^6}{8\pi^3\varepsilon_0^3}
-\frac{8m_0^2E_ne^2\hbar^2}{\pi\varepsilon_0}]\frac{2}{a_0^3n^3l(l+1)(2l+1)}\nonumber \\
&-&\frac{9m_0^2e^4\hbar^2}{4\pi^2\varepsilon_0^2}[\frac{12}{a_0^4n^3l(l+1)(2l+3)(4l^2-1)}
-\frac{4}{a_0^4n^5(2l+3)(4l^2-1)}]\nonumber \\
&+&\frac{m_0e^2\hbar^2}{\pi\varepsilon_0}[6l(l+1)\hbar^2
-12\hbar^2][\frac{20}{a_0^5n^3l(l^2-1)(l+2)(2l+3)(4l^2-1)}\nonumber \\
&-&\frac{20}{3a_0^5n^5(l-1)(l+2)(2l+3)(4l^2-1)}-\frac{4}{3a_0^5n^5l(l^2-1)(l+2)(2l+1)}].
\end{eqnarray}
Using he expectation value of $p^2$, $p^4$ and $p^6$ we finally
land on to the required expression of the  modified eigenvalues
for Hydrogen atom:
\begin{eqnarray}
{E_{n}}^{(mha)}=& & E_n+m_0+\frac{\alpha
E_n^4}{2m_0(m_P)^2}-2m_0[E_n+\frac{\hbar^2}{m_0a_0^2n^2}](\frac{\alpha
E_n^4}{4m_0^3(m_P)^2}+\frac{\alpha E_n^2}{m_0(m_P)^2})\nonumber \\
&+&[\frac{8\hbar^4}{a_0^4n^3(2l+1)}+4m_0^2E_n^2+\frac{8m_0E_n\hbar^2}{n^2a_0^2}]\nonumber \\
& &(\frac{\alpha}{2m_0(m_P)^2}+\frac{\alpha
E_n^2}{2m_0^3(m_P)^2}-\frac{1}{8(m_0)^3} +\frac{3\alpha
E_n^4}{16m_0^5(m_P)^2})\nonumber \\
&+&[8m_0^3E_n^3+\frac{6m_0^3e^2E_n^2}{\pi\varepsilon_0}\frac{1}{n^2a_0}
+\frac{3m_0^3e^4E_n}{2\pi^2\varepsilon^2}\frac{2}{n^3a_0^2(2l+1)}\nonumber \\
&+&(\frac{m_0^3e^6}{8\pi^3\varepsilon_0^3}
-\frac{8m_0^2E_ne^2\hbar^2}{\pi\varepsilon_0})\frac{2}{a_0^3n^3l(l+1)(2l+1)}\nonumber \\
&-&\frac{9m_0^2e^4\hbar^2}{4\pi^2\varepsilon_0^2}(\frac{12}{a_0^4n^3l(l+1)(2l+3)(4l^2-1)}
-\frac{4}{a_0^4n^5(2l+3)(4l^2-1)})\nonumber \\
&+&\frac{m_0e^2\hbar^2}{\pi\varepsilon_0}(6l(l+1)\hbar^2-12\hbar^2)
(\frac{20}{a_0^5n^3l(l^2-1)(l+2)(2l+3)(4l^2-1)}\nonumber \\
&-&\frac{20}{3a_0^5n^5(l-1)(l+2)(2l+3)(4l^2-1)}-\frac{4}{3a_0^5n^5l(l^2-1)(l+2)(2l+1)})]
\times \nonumber \\
&&(-\frac{\alpha}{4m_0^3(m_P)^2}+\frac{1}{16(m_0)^5}-\frac{3\alpha
E_n^2}{8m_0^5(m_P)^2}-\frac{5\alpha
E_n^4}{32m_0^7(m_P)^2}).\label{LSHA1}
\end{eqnarray}
The expression of modified eigenvalues obtained in (\ref{LSHA1})
is identical to the expression as we have already obtained in
(\ref{LSHA}). Thus the result that has been obtained using
Schrodinger equation agrees with the result obtained using
differential form of operator $p^2$. Like the harmonic oscillator
each level of hydrogen atom too has acquired Planck-scale
correction due to the strong the gravity background at that energy
scale. In the article \cite{PSR1}, too we  fond that all the
levels of Hydrogen atom got Planck-scale correction. Albeit the
expression of corrected eigenvalues are different in these two
cases it as expected to have agreement of the correction
numerically with suitable settings of the parameters.  It is true
that the correction due to this generalized MDR is very small and
it may be the case that it is beyond the scope of experimental
verification even with the hitherto available advanced
instrumental facilities. But from the theoretician point of view
this correction cannot be ignored at the present state of time
when interest towards obtaining Planck-scale correction has been
getting intensified. It is worth mentionable that the expression
obtained in (\ref{LSHA}), and (\ref{LSHA1}) are not Lorentz
symmetric because we have not considered the full relativistic
theory since analytical extension with full relativistic structure
is not tractable with this framework.

\section{{\bf Discussion and conclusion}}
We have put forwarded a novel generalized modified dispersion
relation which is expected to be equally useful in order to
incorporate the Planck-scale correction in any  physical system.
The interesting aspect of this novel MDR is that one can use it
maintaining the Lorentz covariance as well as without  the
maintenance of it. In fact, it is a specific choice of the
function $g(\frac{p}{E})$ which keeps the Lorentz symmetry intact,
else it violates Lorentz symmetry.

When the function $g(\frac{p}{E})$ be Lorentz symmetric it can be
considered as physics of modified $d'$ Alambertian $(f(\Box)$ as
introduced in \cite{GAC}. Here modified part will render the
Plank-scale effect within the system. For all energy scale this
MDR will be applicable: below Plank-scale the usual part will be
useful to describe the dynamics, but in the vicinity of the
Plank-scale the modified dispersion will sere the purpose. A
question of positive definite ness may arise,  but  it will not
create any problem here since the original contribution is
dominating so far construction is concerned.

We have  studied the toy model like relativistic harmonic
oscillator with this  MDR with a specific Lorentz symmetric
structure. To get first order correction one needs to compute the
average values up to sixth power of momentum. The spectrum is
determined both by the use of rasing and lowering operator and by
the direct computation using the coordinate representation of the
momentum operator. The results comes out in agreement with each
other that must be the case indeed. This Planck-scale corrected
spectrum of the harmonic oscillator would be useful in the study
of oscillating modes in the early stage of evolution of the
Universe when gravity was expedited to be very strong. It has
already been mentioned that the modified spectrum of harmonic
oscillator is in good agreement with spectrum obtained  in the
article \cite{KEMPF}. So our result also gives a message that MDR
and GUP essentially serve the same purpose towards having the
information of the physics at the vicinity of Planck-scale in
indirect manner.

The relativistic Hydrogen atom problem has also been studied with
this MDR having Lorentz covariant structure. Our investigation
with this MDR reveals that like the harmonic oscillator each
levels of Hydrogen atom too gets the Planck-scale correction. Our
endeavor suggests that the Hydrogen atom spectrum certainly had
had the background effect (Quantum Gravity) which was expected to
play prominent role  at the vicinity of  Planck-scale at the time
of initial formation of it when the Universe was in infant stage.
It may help us to have the information of the process of evolution
of the Universe through the spectrum of the Hydrogen atom because
at the time when Hydrogen atom was formed initially it certainly
encountered prominent quantum gravity effect. So our result may
shade light on the evolution history of the Universe from the
study of Hydrogen or Hydrogen like atom.

This novel MDR may be useful to study the Planck-scale effect in
any physical system. The gravity background where ever was
prominent (in the vicinity of Planck-scale), i.e when it play its
role with the strength comparable to electromagnetic background
this novel MDR is expected to render its important service to
provide the correction required there due to the presence of the
strong gravity background.

The Planck-scale corrected modified eigenvalues for both the
harmonic oscillator and Hydrogen atom would have Lorentz covariant
shape for the specific MDR used here for computation of
eigenvalues. However, it is not the case, since the full
relativistic Hamiltonian is not considered here. It is fair to
admit that complete analytical solution for eigenvalues with full
relativistic Hamiltonian for the harmonic oscillator or Hydrogen
atom with this MDR is not tractable too. However if it would be
possible to proceed with the full complicated structure that would
certainly be much involved and it would exhibit Lorentz symmetric
expression.

 To tell about the future prospect of this novel MDR we would like to add
 that the generalized uncertainty relation corresponding to
 this MDR is indeed a matter of further investigation. It may
be instructive to obtain Planck-scale correction with the use of
this MDR where ever that correction is needed . The extension of
black-hole physics with this MDR would be of interest.
\section{Acknowledgement}
AR likes to thanks E. C. Vagenas for a a helpful discussion at the
early stages of this work.  He acknowledges the facilities
extended to him during his visit to the I.U.C.A.A, Pune. He also
likes to thank the Director of Saha Institute of Nuclear Physics,
Kolkata, for providing library facilities of the Institute.

\end{document}